\documentclass[12pt]{article}

\usepackage[utf8]{inputenc} % set input encoding (not needed with XeLaTeX)

\usepackage{geometry} % to change the page dimensions
\usepackage{graphicx} % support the \includegraphics command and options
\usepackage{booktabs} % for much better looking tables
\usepackage{array} % for better arrays (eg matrices) in maths
\usepackage{paralist} % very flexible & customisable lists (eg. enumerate/itemize, etc.)
\usepackage{verbatim} % adds environment for commenting out blocks of text & for better verbatim
\usepackage{amsbsy}
\usepackage{amsmath}
\usepackage{longtable}
\usepackage[toc,page]{appendix}
\usepackage{natbib}
\usepackage{url}
\RequirePackage[colorlinks,citecolor=blue,urlcolor=blue]{hyperref}
\usepackage{breakurl}
\usepackage{amssymb}

\usepackage[textsize=small]{todonotes}
\usepackage{subcaption}

\usepackage{multirow}

\usepackage{dcolumn}
\raggedright
\graphicspath{{figures/}}

\renewcommand{\tableofcontents}{}

\bibpunct{(}{)}{;}{a}{}{,}  % this is a citation format command for natbib

\begin{document}
%\begin{flushright}
%Version dated: \today
%\end{flushright}

\bigskip
\medskip
\begin{center}

\noindent{\Large \bf Online Bayesian phylodynamic inference in BEAST with application to epidemic reconstruction} \\
%\noindent{\Large \bf to identify site- and branch-specific} \\
%\noindent{\Large \bf evolutionary variation}

\bigskip

\noindent{\normalsize \sc
	Mandev S.~Gill$^{1}$,
	Philippe Lemey$^1$,
	Marc A.~Suchard$^{2,3,4}$,
	Andrew Rambaut$^{5,6}$, and
	Guy Baele$^{\ast,1}$}\\
%\noindent {\small
	\it $^1$Department of Microbiology, Immunology and Transplantation, Rega Institute, KU Leuven, Leuven, Belgium\\
	\it $^2$Department of Human Genetics, David Geffen School of Medicine, University of California, Los Angeles, CA 90095, USA\\
	\it $^3$Department of Biostatistics, School of Public Health, University of California, Los Angeles, CA 90095, USA\\
	\it $^4$Department of Biomathematics, David Geffen School of Medicine, University of California, Los Angeles, CA 90095, USA\\
	\it $^5$Institute of Evolutionary Biology, University of Edinburgh, King's Buildings, EH9 3FL, UK\\
	\it $^6$Fogarty International Center, National Institutes of Health, Bethesda, MD, USA
	
\end{center}
\medskip
\noindent{\bf Corresponding author:} Guy Baele, Department of Microbiology, Immunology and Transplantation, Rega Institute, KU Leuven, Herestraat 49, 3000 Leuven, Belgium; email: \url{guy.baele@kuleuven.be}\\

%\vspace{1in}

%\clearpage

\newcommand{\psaB}{\emph{psaB }}
\newcommand{\ndhD}{\emph{ndhD }}

\newcommand{\numTaxa}{N}
\newcommand{\numColumns}{C}
\newcommand{\columnIdx}{c}
\newcommand{\branchIdx}{b}
\newcommand{\branchLength}[1]{t_{#1}}
\newcommand{\ctmcProbability}[3]{\mathbb{P}_{#1 #2}( #3 )}
\newcommand{\asrvMultiplier}[1]{r_{#1}}
\newcommand{\branchMultiplier}[1]{\mu_{#1}}

\newcommand{\order}[1]{{\cal O} \left( #1 \right)}

\newcommand{\numStates}{S}
\newcommand{\stateIdx}{s}
\newcommand{\stateIdxTwo}{\stateIdx^{\prime}}

\newcommand{\numModels}{K}
\newcommand{\modelIdx}{k}
\newcommand{\modelIdxTwo}{\modelIdx^{\prime}}

\newcommand{\mmRateMatrix}{\mathbf{\Lambda}}
\newcommand{\rateMatrix}[1]{\mathbf{Q}_{#1}}
\newcommand{\probMatrix}[1]{\mathbf{P}_{#1}}
\newcommand{\rateMatrixElement}[3]{Q^{(#1)}_{#2 #3}}
\newcommand{\stationaryDistribution}[2]{\pi_{#1 #2}}

\newcommand{\switchingRate}[2]{\phi_{#1 #2}}
\newcommand{\identityMatrix}[1]{\mathbf{I}_{#1}}
\newcommand{\switchingMatrix}[2]{\switchingRate{#1}{#2} \identityMatrix{}}
\newcommand{\rateMultiplier}[1]{\rho_{#1}}
\newcommand{\scaledRateMatrix}[1]{\rateMultiplier{#1} \rateMatrix{#1}}
\newcommand{\tm}{\text{-}}
\newcommand{\regimeMatrix}{\mathbf{\Phi}}

\newcommand{\diagonalRateMatrix}[1]{\coloredBox{\scaledRateMatrix{#1}}
- \sum\limits_{\modelIdx \neq #1} \switchingMatrix{#1}{\modelIdx}
}

\newcommand{\bigzero}[2]{\makebox(0,0){\hspace{#1}\vspace{#2}\text{\huge0}}}

\newcommand{\coloredBox}[1]{%
\raisebox{-0.5em}{
\tikz \node[
		fill=orange!20,
		align=center,
		minimum width={%
		%5em
		3em
		}
	] (0,0) {%
\ensuremath{#1}\hspace{-0.5em} %
}; %
} % raisebox
}

\paragraph{Abstract}

Reconstructing pathogen dynamics from genetic data as they become available during an outbreak or epidemic represents an important statistical scenario in which observations arrive sequentially in time and one is interested in performing inference in an `online' fashion.
Widely-used Bayesian phylogenetic inference packages are not set up for this purpose, generally requiring one to recompute trees and evolutionary model parameters \emph{de novo} when new data arrive.
To accommodate increasing data flow in a Bayesian phylogenetic framework, we introduce a methodology to efficiently update the posterior distribution with newly available genetic data.
Our procedure is implemented in the BEAST 1.10 software package, and relies on a distance-based measure to insert new taxa into the current estimate of the phylogeny and imputes plausible values for new model parameters to accommodate growing dimensionality.
This augmentation creates informed starting values and re-uses optimally tuned transition kernels for posterior exploration of growing data sets, reducing the time necessary to converge to target posterior distributions.
We apply our framework to data from the recent West African Ebola virus epidemic and demonstrate a considerable reduction in time required to obtain posterior estimates at different time points of the outbreak.
Beyond epidemic monitoring, this framework easily finds other applications within the phylogenetics community, where changes in the data -- in terms of alignment changes, sequence addition or removal -- present common scenarios that can benefit from online inference.

\clearpage

\section{Introduction}
\label{sec:intro}

Changes in data during ongoing research commonly occur in many fields of research, including phylogenetics.
These typically include the addition of new sequences as they become available -- for example, during a large sequencing study or through data sharing -- and updates of alignments of existing sequences, possibly as a result of correcting sequencing errors.
Such changes usually lead to the discarding of results obtained prior to the revision of the data, and recommencing statistical analyses completely from scratch (\emph{de novo}).
Bayesian phylogenetic inference of large data sets can be very time consuming, sometimes requiring weeks of computing time, even when using state-of-the-art hardware.
A promising avenue to mitigate this problem is an online phylogenetic inference framework that can accommodate data changes in existing analyses and leverage intermediate results to shorten the run times of updated inferences.
\par
Existing methods to update phylogenetic estimates in an online fashion are limited, but the initial concept dates back to seminal work by \citet{Felsenstein1981}, who proposed sequential addition of species to a topology as an effective search strategy in tree space.
The stepwise addition approach inserts a new taxon on the branch of the tree that yields the highest likelihood \citep{Felsenstein1993}, and was among the first heuristics to search for a maximum likelihood tree topology.
This concept has also been incorporated into the design of various tree transition kernels and estimation heuristics.
For example, in searching for the optimal tree topology in a maximum-likelihood framework, \citet{Whelan2007} proposed to first \emph{pluck} a number of sequences from an existing tree and subsequently place each sequence onto the tree where it yields the highest likelihood value.
\par
Initial developments to update phylogenies with new sequence data focused on methods for phylogenetic placement, where unknown query sequences -- typically short reads obtained from next-generation sequencing -- are placed onto a fixed tree pre-computed from a reference alignment.
Employing a likelihood-based approach, \citet{Matsen2010} proposed a two-stage search algorithm to accelerate placements for query sequences, where a quick first evaluation of the tree is followed by a more detailed search in high-scoring parts of the tree.
An increasing body of work mainly targets such taxonomic identification methods, with recent developments confronting the increasing scalability issues associated with the high dimensions of modern data sets \citep{Barbera2018,Czech2018}.
\par
\citet{Izquierdo-Carrasco2014} implemented an online framework to estimate phylogenetic trees using maximum-likelihood heuristics, which automatically extends an existing alignment when sufficiently new data have been generated and subsequently reconstructs extended phylogenetic trees by using previously inferred smaller trees as starting topologies.
The authors compared their methodology to \emph{de novo} phylogenetic reconstruction and found a slight but consistent
improvement in computational performance and a similar topological accuracy.
\par
Recent foundational work towards online Bayesian phylogenetic inference focuses on sequential Monte Carlo (SMC) methods to update the posterior distribution \citep{Everitt2016, Fourment2018, Dinh2018}.
These methods approximate a posterior distribution using a set of particles that exist simultaneously, which are updated when new data arrive
and are then resampled with weights determined by the unnormalized posterior density \citep{DoucetSMC}.
While SMC methods are not new to Bayesian phylogenetics, they have primarily been explored to increase computational efficiency in standard inference, for example, to infer rooted, ultrametric \citep{Bouchard2012} and non-ultrametric phylogenetic trees \citep{wang2015bayesian, wang2018annealed}.
Within an SMC framework, \citet{Everitt2016} introduced the use of deterministic transformations to move particles effectively between target distributions with different dimensions and applied this methodology to infer an ultrametric phylogeny of a bacterial population from DNA sequence data.
A similar methodology was developed independently and almost simultaneously by \citet{Dinh2018}, who also describe important theoretical results on the consistency and stability of SMC for online Bayesian phylogenetic inference.
Building upon the work of \citet{Dinh2018}, \citet{Fourment2018} showed that the total time to compute a series of unrooted phylogenetic trees as new sequence data arrive can be reduced significantly by proposing new phylogenies through \emph{guided} proposals that attempt to match the proposal density to the posterior.
All of these SMC approaches focus on the tree inference problem rather than the estimation of broader phylogenetic models where the goal is to marginalize these over plausible trees.
They have also not yet led to implementations in widely-used software packages for Bayesian phylogenetic inference.
\par
The need for online phylogenetic inference is especially pressing in the growing field of phylodynamics (see e.g.~\citet{Baele2016,Baele2018} for an overview).
Phylodynamic inference has emerged as an invaluable tool to understand outbreaks and epidemics \citep{Pybus2012, Faria2014, Worobey2014, Nelson2015, Dudas2017, Metsky2017}, and has the potential to inform effective control and intervention strategies \citep{Dellicour2018, Al-Qahtani2017}.
Importantly, phylodynamic analyses of pathogen genome sequences sampled over time reveal events and processes that shape epidemic dynamics that are unobserved and not obtainable through any other methods.
The Bayesian Evolutionary Analysis by Sampling Trees (BEAST) version 1 software package \citep{Suchard2018} has become a primary tool for Bayesian phylodynamic inference from genetic sequence data, offering a wide range of coalescent, trait evolution and molecular clock models to study the evolution and spread of pathogens, as well as potential predictors for these processes.
\par
Recent advances in portable sequencing technology have led to a reduction in sequencing time and costs, enabling in-field sequencing and real-time genomic surveillance as an outbreak unfolds.
This was demonstrated during the recent Ebola epidemic in West Africa \citep{Quick2016, Arias2016}, as well as  the recent Zika outbreak in the Americas \citep{Faria2017}.
Notably, \citet{Quick2016} were routinely able to sequence Ebola-positive samples within days of collection, and in some cases were able to obtain results within 24 hours.
Such a continuous stream of new sequence data creates the potential for phylodynamic inference to take up a more prominent role in the public health response by providing up-to-date, actionable epidemiological and evolutionary insights during the course of an ongoing outbreak.
Bayesian modeling naturally accommodates uncertainty in the phylogeny and evolutionary model parameters, and therefore offers a coherent inference framework for relatively short outbreak timescales for which the phylogeny may not be well-resolved.
\par
However, the potential of phylodynamic methods in real-time epidemic response can only be fully realized if accurate up-to-date inferences are delivered in a timely manner.
Fast maximum likelihood-based methods, such as those adopted by Nextstrain \citep{Hadfield2018},
can provide rapid updates by relying on a pipeline of fast, but less rigorous heuristic methods \citep{Sagulenko2018}.
Bayesian phylodynamic models rely on MCMC estimation procedures that can have very long run times, often requiring days or weeks to infer the posterior distribution for complex models.
Having to restart these time-consuming procedures when new data become available thus represents a significant impediment to providing regular, updated phylodynamic inferences.
\par
Here we explore an approach that is conceptually simpler than SMC and consists of interrupting an ongoing MCMC analysis upon the arrival of new sequence data and after the current analysis has converged, placing the new sequences at plausible locations in the current tree estimate, and then resuming the analysis with the expanded data set.
We apply this methodology to data from several time periods throughout the West African Ebola virus epidemic of 2013-2016 and show that resuming an interrupted analysis after inserting new sequences into the current tree estimate, as opposed to restarting from scratch, reduces the time necessary to converge to the posterior distribution.
Specifically, our approach virtually eliminates the MCMC burn-in when computing updated inferences that incorporate new data sequenced during a subsequent epidemiological week (epi week, labeled 1 to 52).
This improved efficiency will allow the analysis and interpretation to more closely maintain a real-time relationship to the accumulation of data.

\section{New Approaches}

We present an online phylogenetic inference framework, implemented in the BEAST 1.10 software package, that allows incorporating new data into an ongoing analysis.
Notably, this methodology efficiently updates the posterior distribution upon the arrival of new data by using previous inferences to minimize the burn-in time (the time necessary for the MCMC algorithm to converge to the posterior distribution) for analysis of the expanded data set that includes the new data (along with the previously available data).
Additionally, our implementation includes a new feature for BEAST 1.10 that enables resuming an MCMC analysis from the iteration at which it was terminated (similar to the ``stoppb'' feature in the Bayesian phylogenetics package PhyloBayes \citep{Lartillot2009} ).
\par
When new sequence data become available and the current BEAST analysis has converged to the target distribution, the BEAST analysis is interrupted and a draw (featuring estimates of all model parameters) is taken from its posterior sample.
We insert the new sequences into the phylogenetic tree estimate obtained from the draw in a stepwise fashion, where the location of each insertion is determined by computing the genetic distance between the new sequence and the taxa in the tree.
Next, we impute plausible values for new model parameters that are necessitated by the increased dimensionality of the enlarged phylogenetic tree, such as branch-specific evolutionary rates.
Parameter values for models unaffected by the increased data dimensionality are left unchanged.
The BEAST analysis is then resumed with the simulation of an MCMC sample with starting parameter values that have been constructed from the aforementioned imputation and sequence insertion algorithm.
Further, the resumed analysis employs a stored set of MCMC transition kernels that have been optimized for efficient sampling using BEAST's auto-tuning functionality.
\par
To determine the performance of this framework, we carefully assess the reduction in time required to converge to the target posterior distribution by using both visual analyses of MCMC trace plots as well as a scripted sliding window approach to determine burn-in.
The various steps of this approach are described in more detail in Materials and Methods.
We provide BEAST XML input files for the analyses performed throughout this paper as well as a tutorial on setting up these analyses at \url{http://beast.community/online_inference.html}.
The tutorial also describes how to set up an MCMC analysis so that it can be resumed from the iteration at which it was terminated. This new feature in BEAST 1.10 will be useful in general (beyond an online inference setting), for example, in the case of a computer crash, or if an MCMC analysis needs to be run for longer to generate sufficient samples.

\section{Results}
\label{sec:results}

We evaluate the performance of our BEAST 1.10 online inference framework by analyzing complete genome data from the West African Ebola virus epidemic of 2013-2016.
The data comprise 1610 whole genome sequences collected throughout the epidemic, from 17 March 2014 to 24 October 2015 \citep{Dudas2017}.
Each sequence is associated with a particular epi week during which the sample was obtained, allowing us to recreate a detailed data flow of the actual epidemic.
For the purpose of our performance comparisons, we assume that the genome data were made available immediately after the time of sampling, allowing us to assess potential efficiency gains in a scenario where a Bayesian phylodynamic reconstruction would be attempted once per epi week, incorporating the newly obtained genome data into the inference up to the previous epi week.
\par
Although our previous study on these data was performed towards the end of the epidemic \citep{Dudas2017}, during this work we were still confronted with new genome sequences becoming available, requiring us to frequently restart our MCMC analyses \emph{de novo}.
Considering the size of the data set, this required tremendous computational effort to obtain updated results.
Here, we evaluate our online procedure by computing updated inferences corresponding to increases in data during consecutive epi weeks at different time points during the epidemic.
For each time point we consider two consecutive epi weeks, which we shall refer to as the \textit{first} and \textit{second} epi weeks in this context.
We analyze the cumulative data available by the end of the second epi week using two methods: our proposed online inference framework which augments a previous analysis with newly obtained data (see Materials and Methods), and a \emph{de novo} analysis using a randomly generated starting tree and default starting values for the model parameters following a typical Bayesian phylogenetic analysis.
We use a slightly different phylodynamic model setup than in our previous study \citep{Dudas2017}, i.e.~an exponential growth coalescent model as the prior density on trees \citep{Griffiths1994}, and an HKY$+\Gamma4$ substitution model \citep{HKY,Yang1996} for each of the four nucleotide partitions (the three codon positions and the non-coding intergenic regions) with different relative rates across the partitions.
Evolutionary rates were allowed to vary across branches according to an uncorrelated relaxed molecular clock model with an underlying log-normal distribution \citep{Drummond06}.
The overall evolutionary rate was given an uninformative continuous-time Markov chain (CTMC) reference prior \citep{ferreira08}, while the rate multipliers for each partition were given a joint Dirichlet prior.
The BEAST 1.10 XML files used in our analyses are available at \url{http://beast.community/online_inference.html}.
\par
\begin{figure*}[h]
\begin{center}
\includegraphics[width=1.0\textwidth]{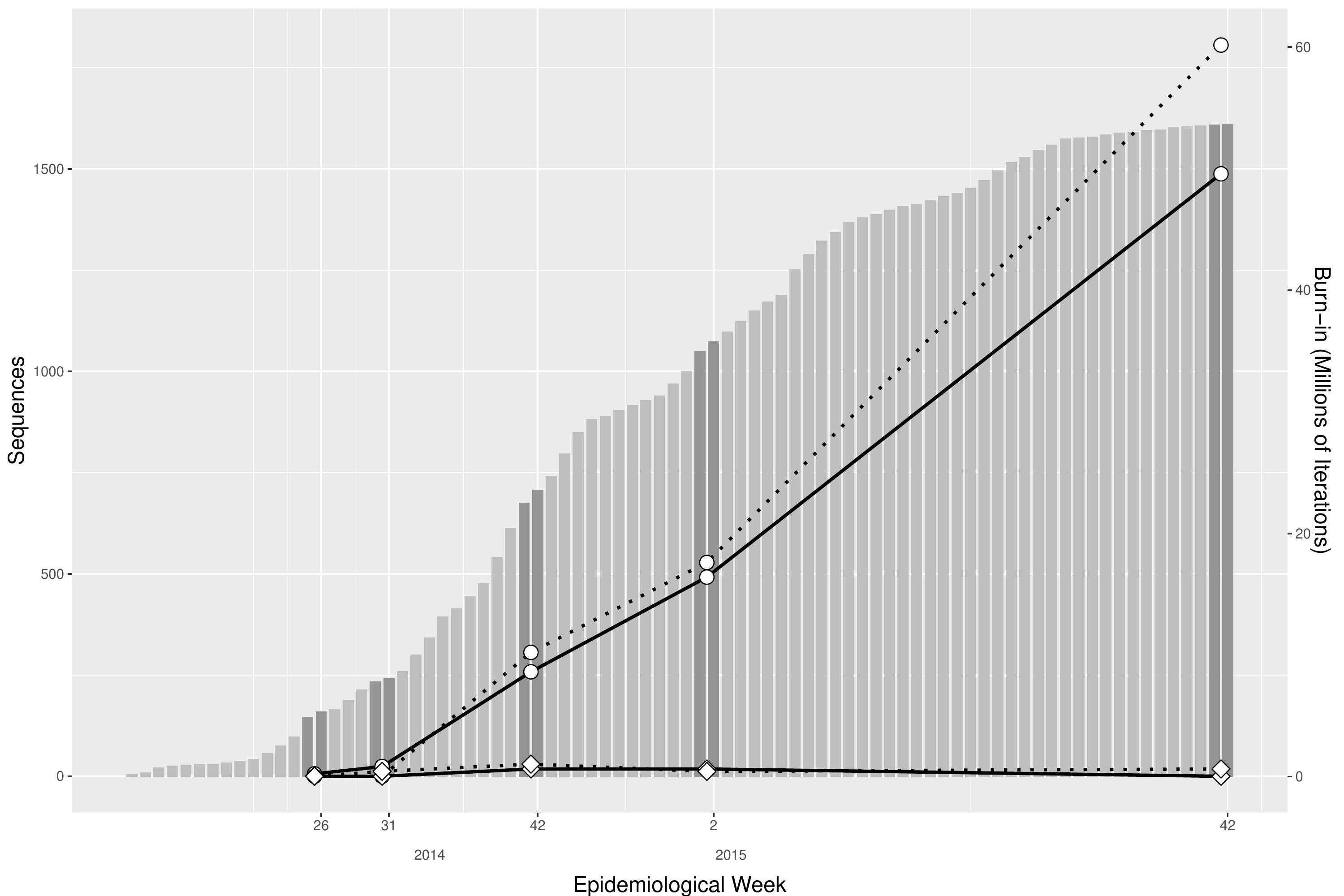}
\end{center}
\caption[Reduced burn-in resulting from online Bayesian analyses compared to standard \emph{de novo} analyses.]
{Comparison of burn-in resulting from standard \emph{de novo} analyses versus online Bayesian analyses to compute updated inferences from data taken from different time points of the West African Ebola virus epidemic. 
The data flow of the epidemic, in terms of total sequence available during each epi week, is recreated in the background of the plot in gray bars.
Dark gray bars show the data corresponding to the five time points at which we compute updated inferences.
The plots chart the burn-in required by \emph{de novo} analyses, represented by circles, and online analyses, represented by diamonds.
Solid lines correspond to burn-in estimates based on visual analyses of trace plots while dotted lines correspond to burn-in estimates based on maximizing ESS values.
}
\label{fig:barplot}
\end{figure*}

We consider five different pairs of consecutive epi weeks from the 2013-2016 Ebola epidemic: epi weeks 25 and 26 of 2014, epi weeks 30 and 31 of 2014, epi weeks 41 and 42 of 2014, epi weeks 1 and 2 of 2015, and the final epi weeks 41 and 42 of 2015.
These sets of epi weeks constitute a relatively broad range of possible sequence addition scenarios, as they occurred during the actual epidemic.
We provide details on the number of sequences for these scenarios in Table~\ref{tab:burnin} and Figure~\ref{fig:barplot}.
As a Markov chain constitutes a stochastic process, for each time point we perform five independent replicates of a standard \emph{de novo} analysis of the data available by the end of first epi week, five independent replicates of a standard \emph{de novo} analysis of the data available by the end of the second epi week, and five independent replicates of an online analysis of the data available by the end of the second epi week. Note that each online analysis proceeds by updating inferences from one of the \emph{de novo} analyses of the data available by the end of the first epi week.
We examine split frequencies for tree samples from independent replicates to compare replicates and ensure convergence to the same posterior distribution (see Supplementary Material). In particular, in all analyses we observe an average standard deviation of split frequencies that meets the guideline of being less than 0.01 (see Materials and Methods). The replicates are independent in that the MCMC simulations start from different trees.  In particular, standard \emph{de novo} analyses use randomly-generated starting trees, and online analyses feature starting trees that differ because they are constructed by augmenting different tree estimates from different \emph{de novo} analyses of the data available by the end of the first epi week.
For each time period, we determine a random order for the new sequences and insert them into the tree estimate in the same order for each of the five replicates.

\begin{table*}[h]
\begin{center}
\begin{tabular}{lcccccc}
\cline{1-7}
 &\multicolumn{2}{c}{Sequences} & \multicolumn{2}{c}{Standard analysis} & \multicolumn{2}{c}{Online analysis} \\
 \multicolumn{1}{c}{Data} & \multicolumn{1}{c}{Total} & \multicolumn{1}{c}{Added}
 & \multicolumn{1}{c}{Burn-in (G)} & \multicolumn{1}{c}{Burn-in (ESS)}
& \multicolumn{1}{c}{Burn-in (G)} & \multicolumn{1}{c}{Burn-in (ESS)} \\
\hline
2014, Epi week 26 & 158 & 13 & 0.2 ($<$0.1) & $<$0.1 ($<$0.1) & $<$0.1 ($<$0.1) & $<$0.1 ($<$0.1)  \\
2014, Epi week 31 & 240 & 8 & 0.8 (0.3)& $<$0.1 ($<$0.1) & $<$0.1 ($<$0.1) & 0.4 (0.9) \\
2014, Epi week 42 & 706 & 32 & 8.6 (2.1) & 10.2 (10.3) & 0.6 (0.9) &  1.0 (1.0)   \\
2015, Epi week 2 & 1072 & 24 & 16.4 (7.3) & 17.6 (7.1) & 0.6 (0.5) & 0.4 (0.5) \\
2015, Epi week 42 & 1610 & 2 & 49.6 (20.6) & 60.2 (15.4) & $<$0.1 ($<$0.1) & 0.6 (1.3)  \\
\hline
\end{tabular}
\end{center}
\caption[Reduced burn-in achieved with online Bayesian phylodynamic inference.]
{Reduced burn-in (in millions of iterations) achieved with online Bayesian phylodynamic inference.
Comparison of burn-in for the log joint density sample resulting from two different analysis methods applied to Ebola virus data taken from the West African Ebola epidemic of 2013-2016.
The standard \emph{de novo} approach of analyzing the full data set from scratch is compared to the online inference approach that updates inferences from the previous epi week upon the arrival of new data.
The length of burn-in (in millions of states) is determined through a graphical approach (G) that consists of analyzing posterior trace plots, as well as by computing the amount of discarded burn-in that maximizes the effective sample size (ESS).
Results are averaged over five replicates for each analysis, with standard deviation in parentheses.
\label{tab:burnin}}
\end{table*}

\par
For each pair of consecutive epi weeks, we compare the burn-in for the sample of the log joint density (which is proportional to the posterior density) resulting from online and standard \emph{de novo} analyses. Figure~\ref{fig:barplot} and Table~\ref{tab:burnin} show the results, averaged over five replicates.
The different methods of determining the burn-in (see Materials and Methods) yield very similar estimates.
We assess the sensitivity of sequence insertion order by performing five additional replicates each for epi weeks 41 and 42 of 2014 and epi weeks 1 and 2 of 2015. Each of the additional replicates for a given time period augments the same inferences through a different, random sequence insertion order. We find that the estimated burn-in for each additional replicate is in line with the burn-in estimate for the corresponding time period in Table~\ref{tab:burnin}, lying within two standard deviations of the mean. 
\par
The results show that our online inference framework can reduce burn-in by a significant amount ($p$-values are less than 0.01 for $t$-tests comparing burn-in from online and standard analyses for the latter three epi weeks).
While the burn-in for epi weeks 26 and 31 of 2014 is negligible in both online and standard analyses, the standard approach requires substantial burn-in in the latter three cases.
By reducing the average burn-in to one million iterations or less for each of these three epi weeks, the online approach virtually eliminates the burn-in in these analyses.
The results for epi week 42 of 2015 data are particularly remarkable (see Figures S1, S2 and S3 for a comparison of posterior trace plots from five replicates of all test cases), showing average reductions of burn-in by 50 to 60 million iterations.

\begin{figure*}[!htbp]
\begin{center}
\includegraphics[width=1.0\textwidth]{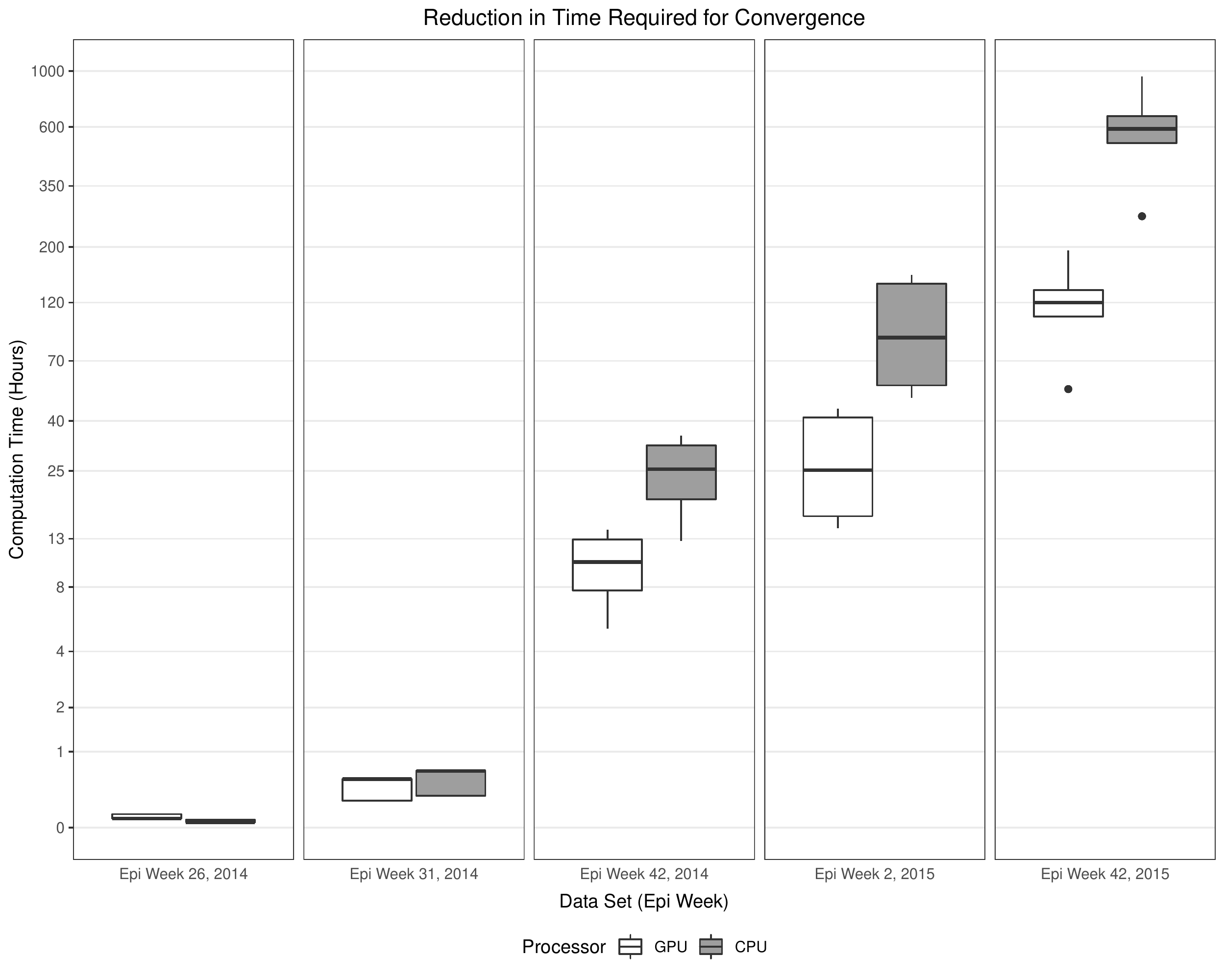}
\end{center}
\caption[Reduction in computation time gained from online inference]
{Box plots show distribution of savings in computation time by using online inference as compared to standard \emph{de novo} analyses to update inferences for data from different time points in the West African Ebola virus epidemic.
White box plots correspond to analyses using a Tesla P100 graphics card for scientific computing and gray boxes correspond to analyses using a multi-core CPU.
Irrespective of the actual hardware used, the time savings are substantial with up to 600 hours on average saved using our online approach on CPU for our most demanding scenario.
The axis corresponding to running time (in hours) is log-transformed to allow for greater visibility of plots for smaller data sets.
}
\label{fig:comptime}
\end{figure*}

\par
To put these efficiency gains into perspective, it is useful to translate the reduction of burn-in into actual saved computing time using a multi-core CPU (in our case, a 14-core 2.20 GHz Intel Xeon Gold 5120 CPU) as well as using a state-of-the-art hardware setup enhanced by a GPU (e.g., a Tesla P100 graphics card intended for scientific computing).
We use BEAGLE 2.1.2 \citep{Ayres2012} to enable such GPU computation within BEAST.
Figure~\ref{fig:comptime} depicts the savings in computation time by using online inference as compared to standard \emph{de novo} analyses to update inferences for data from different time points in the West African Ebola virus epidemic.
Dunn tests \citep{Dunn1961} indicate that the savings under online inference for each time point are significant ($p < 0.01$).
We note that running time depends on burn-in length as well as data set size, with larger data sets requiring more time per iteration.
Our online inference approach leads to higher computation time savings as the complexity of the data increases, with up to 600 hours being saved on average on a modern multi-core processor.
State-of-the-art graphics cards targeting the scientific computing market are able to reduce this number to 120 hours on average of savings, but such cards may not be readily available, especially in resource-limited settings.

\section{Discussion}
\label{sec:discussion}

We present a framework for online Bayesian phylodynamic inference that accommodates a continuous data flow, as exemplified by an epidemic scenario where continued sampling efforts yield a series of genome sequences over time.
This framework has been implemented in BEAST 1.10, a popular software package for Bayesian phylogenetic and phylodynamic inference.
Through empirical examples taken from the 2013-2016 West African Ebola epidemic, we show that our online approach can significantly reduce burn-in and, consequently, the time necessary to generate sufficient samples from the posterior distribution of a phylodynamic model being applied to a growing data set.
The savings in computation time can amount to days or even weeks, depending on the computational infrastructure, the complexity of the data and hence also the accompanying phylodynamic model.
\par
The improvements in computational efficiency through minimizing burn-in that we observe are encouraging, but there is a need to continue improving efficiency in multiple directions.
First, alternative sequence insertion and branch rate imputation procedures may yield better performance in certain situations.
\citet{Desper2002}, for instance, employ a minimum evolution criterion for stepwise addition of taxa.
As another example, an insertion procedure that allows new sequences to have insertion times that are deeper than the root of the current tree estimate may be more suitable in the case that new sequences are distantly related to the sequences that already exist in the tree.
Under the current implementation, MCMC transition kernels enable the insertion point of a new sequence to eventually be repositioned deeper than the root of the starting tree. However, allowing a sequence to be directly inserted deeper than the root may save computational time.
\par
Second, even if burn-in is minimized, generating sufficient samples from the Markov chain after it has converged to the posterior distribution can still be very time-consuming.
A popular approach to generate samples more quickly is to run multiple independent chains, starting from different random locations in search space, in parallel and combine the posterior samples.
However, the time saved through such a strategy depends on the burn-in phase, which must elapse for each chain before its samples can be used.
From this perspective, the advances of our online framework are especially important.
Another strategy for more efficient sampling is to evaluate past MCMC performance during pauses to incorporate new data and make informed adjustments prior to resuming the analysis.
For instance, transition kernel weights can be modified to focus on parameters with low ESS values.
Progress can also be made through advances in MCMC sampling that enable more efficient exploration of posterior distributions. Innovative sampling techniques that have already shown promise in the context of phylogenetics and are ripe for further development include adaptive MCMC \citep{Baele2017} and Hamiltonian Monte Carlo \citep{NealHMC, Lan2015,Ji2019}.
Finally, the computational performance will undoubtedly benefit from continued development of high-performance libraries for phylogenetic likelihood calculation \citep{Ayres2019}.
\par
The implementation we present here differs from other recent work on online Bayesian phylogenetic inference, which relies on SMC to update phylogenies \citep{Everitt2016, Fourment2018, Dinh2018}.
While SMC represents a principled approach to infer a distribution of growing dimensions, the SMC-based methods for online Bayesian phylogenetics are limited to inferring phylogenetic trees.
It would be beneficial to integrate SMC algorithms for updating phylogenies with MCMC methods to sample other evolutionary model parameters, and ultimately to implement a complementary online inference framework in BEAST.
Such an implementation would enable direct comparison of the current online framework with SMC-based approaches, allowing researchers to assess the benefits and drawbacks of each approach and helping to streamline future development of online Bayesian phylogenetic inference.
\par
Our development has been primarily motivated by epidemic scenarios that entail a continuous stream of new sequence data becoming available during the course of an outbreak.
In our empirical assessment of the West African Ebola virus epidemic, we have assumed that the genome data were made available close to the time of sampling, which represents the ideal scenario in an outbreak response.
In reality, during the epidemic, there was considerable variation in how rapidly virus genome data were available for analysis.
There were many reasons for this, but even when genomes were being shared as rapidly as possible, the batch shipping of samples to high-throughput sequencing centers resulted in a minimum delay of many weeks \citep{Gire2014,Park2015}. 
This changed towards the end of the epidemic as new, portable, sequencing instruments were installed in Ebola treatment centers in Guinea and Sierra Leone \citep{Quick2016,Arias2016}, producing virus genome sequences from patients within days or hours of a sample being taken. We expect that the use of such instruments at the point of diagnosis will increase and the resulting stream of sequence data will mean that the computational analysis will become the bottleneck in using the data to inform the response.
From this perspective, the reduction in time necessary to compute updated inferences on data from the Ebola virus epidemic through our online inference framework is promising, and continued efforts to further improve efficiency are crucial.
\par
Beyond computational efficiency, additional development is needed in order to maximize the potential impact of our framework as a support tool during outbreaks.
The current implementation must be extended to accommodate more sophisticated phylodynamic models, especially methods that integrate sequence data with other epidemiological data to elucidate different phylodynamic processes \citep{Lemey2009, Lemey2014, Gill, Gill2016}. For many of these models -- for example, a phylogeographic model for which a sequence from a previously unsampled location is being added -- the addition of novel sequence data will increase their dimensionality, and methods that augment the models in an intelligent manner are essential.
Adding sequence data may also require increasingly complex models to accurately describe the underlying evolutionary processes as the data set grows (e.g.~transitioning from a strict to a relaxed clock model), a process that should ideally not require user interactions. This could potentially be addressed by developing nonparametric Bayesian models for evolutionary heterogeneity that can dynamically accommodate increasing model complexity.
Finally, we have focused on evaluating the performance of updating phylogenetic inferences conditional on pre-aligned sequence data. However, a comprehensive system for real-time evolutionary analysis will need to include an alignment step when new sequence data become available.
\par
Finally, while real-time monitoring of infectious disease outbreaks has motivated much of our development, we anticipate that our online inference framework will be more broadly useful, allowing researchers to save precious time in any context in which new data become available that extend a previously analyzed data set.
Many large-scale sequencing efforts in a wide range of research fields generate a steady flow of genomic data sequences, which often involve a phylogenetic component, and as such online Bayesian phylogenetic inference will prove useful beyond the field of pathogen phylodynamics.

\section{Materials and Methods}
\label{sec:methods}

\subsection{Online Bayesian phylogenetic inference}

Our strategy to increase efficiency through an online inference framework in BEAST 1.10 builds on using estimates from a previous MCMC analysis in order to minimize time to convergence to the new posterior distribution.
In MCMC simulation, this \textit{burn-in} period corresponds to a transient phase of the Markov chain during which the simulated values reflect the influence of the starting values of the chain and are from low-probability regions of the target posterior distribution \citep{Brooks1998}.
The burn-in period ends once the chain achieves stationary behavior and has converged to the posterior distribution.
Including simulated values from the burn-in phase of the chain in approximations of the posterior distribution can lead to substantial bias and it has therefore become common practice to discard samples taken during the burn-in period.
Burn-in phases for standard phylodynamic models on realistic data sets can be extremely long, and through minimizing burn-in, we can save a potentially large proportion of the computational time usually required to generate a good posterior sample.
\par
Online inference can be viewed as a series of steps (or generations) with increasing amounts of data, with each step consisting of sampling from the posterior distribution for the model specified at the given step.
The model must be adjusted when transitioning from one step to the next in order to accommodate the growth in data.
Consider an ongoing (or completed) analysis at step $i$ of a data set of $N_i$ sequences with a phylodynamic model that includes a choice of substitution model(s) \citep{JC69,HKY, Tavare}, a strict or uncorrelated relaxed molecular clock model \citep{Drummond06}, and a parametric coalescent tree prior \citep{Griffiths1994}.
Assume that at step $i$, the analysis has achieved convergence and has generated samples from the posterior distribution.
Upon the arrival of $M_{i+1}$ new sequences, we interrupt the step $i$ analysis (if it has not yet run to completion), augment the analysis with the new sequences, and proceed to step $i+1$, during which we will analyze the expanded data set of $N_{i+1} = N_i+M_{i+1}$ sequences.
\par
\renewcommand{\Theta}{\theta} 
We take a random draw $\boldsymbol \Theta_i$ from the posterior sample (i.e.~excluding the burn-in) generated in step $i$ that
consists of estimates of the phylogenetic tree and all other model parameters.
Further, BEAST automatically optimizes transition kernel tuning parameters during an MCMC analysis in order to maximize sampling efficiency, and we extract the optimized tuning parameter values from step $i$.
We modify the elements of $\boldsymbol \Theta_i$ as necessary in order to obtain $\boldsymbol \Theta_{i+1}^{(0)}$, the starting model parameter
values for the MCMC chain simulated in step $i+1$.  The aim in our construction of $\boldsymbol \Theta_{i+1}^{(0)}$ is to leverage the values of $\boldsymbol \Theta_{i}$ to obtain starting parameter values that are in, or relatively close to, a high-probability region of the target posterior in step $i+1$, and thereby minimize the step $i+1$ burn-in phase.
This is in contrast to the typical approach of using default or randomly generated starting parameter values (including the phylogenetic tree) that can be very distant from high-probability regions of the posterior.
Such suboptimal starting values are a major cause of long burn-in periods.
\par
\begin{figure*}[!htbp]
\begin{center}
\includegraphics[width=1.0\textwidth]{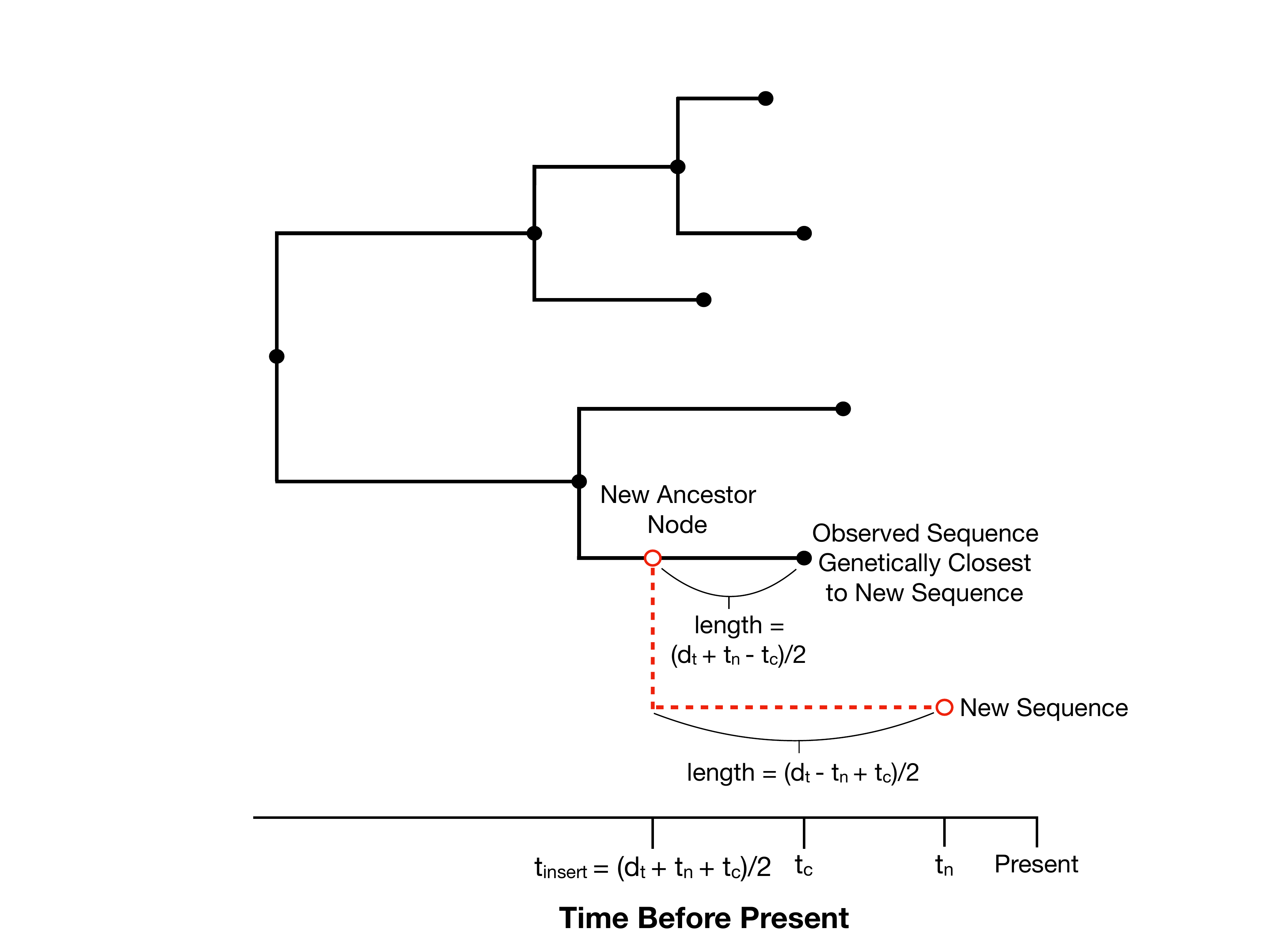}
\end{center}
\caption[Sequence insertion:]
{A new sequence is inserted into an existing phylogenetic tree by determining the closest observed sequence (in terms of genetic distance) already in the tree, and inserting a new ancestor node for the new sequence and its closest sequence. The genetic distance between the new sequence and its closest sequence is converted into a distance in units of time, $d_t$, by dividing by the evolutionary rate associated with the branch leading to the closest sequence. To determine the insertion time $t_{\text{insert}}$ of the new ancestor node (in terms of time prior to the present time), we require $(t_{\text{insert}} - t_c) + (t_{\text{insert}}-t_n) = d_t$, where $t_n$ is the sampling time of the new sequence, and $t_c$ the sampling time of its closest sequence. This yields $t_{\text{insert}} = (d_t + t_n + t_c)/2$.}
\label{fig:insertion}
\end{figure*}

The algorithm to augment $\boldsymbol \theta_i$ to $\boldsymbol \theta_{i+1}^{(0)}$ starts with expanding the tree from $\boldsymbol \theta_i$ by inserting a new sequence into it. The sequence insertion process is illustrated in Figure~\ref{fig:insertion}. First we find the observed sequence already in the tree that is closest to the new sequence in terms of genetic distance, where genetic distance is based on a simple nucleotide substitution model. We compute the genetic distance in all analyses using a JC69 model \citep{JC69}, but our implementation also offers an F84 model \citep{Felsenstein1996}. 
\par
We then insert a common ancestor node for the new sequence and its ``closest'' sequence. To determine the height at which to insert the new ancestor node, we first translate the genetic distance $d$ between the two sequences to a distance $d_t$ in units of time by dividing $d$ by the evolutionary rate associated with the branch leading to the ``closest'' sequence. 
Further, let $t_n$ denote the sampling time (in terms of time units prior to the present time) of the new sequence, $t_c$ the sampling time of the ``closest'' sequence, and $t_{\text{insert}}$ the time at which we will insert the new ancestor node. Assume, without loss of generality, that the new sequence has a more recent sampling time (so that $t_c > t_n$). Consider
\begin{equation}
t^* = t_c + \frac{d_t - (t_c-t_n)}{2} = \frac{d_t + t_c + t_n}{2}.
\label{eq:eq1}
\end{equation}
We set $t_{\text{insert}} = t^*$ (except in special cases, which we discuss shortly) because this ensures that the placement of the new ancestor node is consistent with $d_t$ in that $(t_{\text{insert}} - t_c) + (t_{\text{insert}} - t_n) = d_t$. Notably, this method of determining the insertion height allows the new branch to emanate from an external branch or internal branch, with the latter case accommodating realistic insertion of divergent lineages. In certain cases, however, we use an alternative insertion time because setting $t_{\text{insert}} = t^*$ results in a new branch of length 0 or an insertion time greater than or equal to $t_{\text{root}}$, the root height of the tree. Let $\epsilon$ denote a scalar in the interval $(0,1)$, let $t_{\text{child}}$ denote the height of the child node of the branch that will be split by the insertion of the new ancestor node, and let $l_b$ denote the length of the aforementioned branch. In the case that $t^* \geq t_{\text{root}}$, $t_{\text{child}}$ and $l_b$ correspond to the ancestral branch of the ``closest'' sequence that connects to the root. (If $t^*$ is equal to the height of an internal node, we adopt the convention that the branch for which this internal node is the parent is the branch that will be split.) Then if $t^* \geq t_{\text{root}}$ or if $t^*$ is equal to the height of a node on the branch that will be split, then we set $t_{\text{insert}} = t_{\text{child}} + \epsilon * l_b$. See Algorithm S1 in the Supplementary Material for further details.

\par
Next, the growth of the tree after a sequence insertion requires branch-specific aspects of the evolutionary model to assume a greater dimension.
In particular, our implementation allows for specification of either a strict or uncorrelated relaxed molecular clock model.
Under the uncorrelated relaxed clock, each branch-specific clock rate is drawn independently from an underlying rate distribution (e.g.~an exponential or log-normal distribution).
The underlying rate distribution is discretized into a number of categories equal to the number of branches, and each branch receives a unique clock rate corresponding to its assigned category.
We impute clock rates on the branches of the enlarged tree by assigning branches to rate categories according to a deterministic procedure described in detail in the Supplementary Material.
\par
The algorithm continues in this fashion: the remaining new sequences are inserted into the growing phylogenetic tree one at a time, and uncorrelated relaxed clock rates associated with tree branches are updated after each insertion. 
The order of insertion can be specified by the user in the XML (in the Ebola virus example, a sensitivity analysis detailed in the Results section suggests that the performance does not depend on insertion order). 
Aspects of the model that remain compatible with an increase in sequence data, such as substitution model specification, are left unaltered, and the parameters that characterize these aspects are identical in both $\boldsymbol \Theta_i$ and $\boldsymbol \Theta^{(0)}_{i+1}$.
\par
The final part of step $i+1$ is to simulate a Markov chain, with starting model parameter values $\boldsymbol \Theta^{(0)}_{i+1}$ and initial tuning parameter values taken, pre-optimized, from step $i$.
We note that there is no hard-encoded stopping rule, and the termination of the simulated chain at step $i+1$ is left to the user's discretion.
The simulation should continue at least until the chain has achieved stationarity, and until either new data become available (and the simulation can be interrupted to incorporate the new data), or a sufficient posterior sample for inference has been produced.
However, there is no need to completely terminate the chain at step $i+1$ if it is interrupted to incorporate new data because the step $i+1$ chain can be resumed after the interruption, and the step $i+2$ simulation for the expanded data set can be started as an independent process.
Indeed, if step $i+1$ has yet to produce sufficient posterior samples it may be optimal to resume its simulation to obtain provisional inferences (that could go towards informing the response to an outbreak, for instance) while waiting for the step $i+2$ chain to converge.

\subsection{Performance}
We assess burn-in using two different approaches.
First, we use Tracer \citep{Rambaut2018}, a popular software package for posterior summarization in Bayesian phylogenetics, to visually examine trace plots of the posterior distribution.
The earliest iteration after which the plot exhibits stationarity is taken to be the end of the burn-in period.
Second, we use the R \citep{CRANR} package coda \citep{coda} to compute the effective sample size (ESS) of the log joint (likelihood $\times$ prior) density sample after discarding the first $n$ samples, and we adopt the value of $n$ that yields the maximal ESS as the burn-in.
The ESS is a statistic that estimates the number of independent draws from the target distribution that an MCMC sample corresponds to by accounting for the autocorrelation in the sample \citep{Kass1998}, and the joint density is often, even by us, called the ``posterior'' in BEAST.
This is inexact because the joint density is an unnormalized rescaling of the posterior.
Discarding highly correlated burn-in iterates from the sample leads to a greater ESS and, in effect, a more informative sample.
\par
We compare the frequencies of splits (or clades) across multiple independent Markov chains in order to ensure that the independent replicates for a given time point in the Ebola virus epidemic converge to the same stationary distribution. In particular, we compare chains generated by the same method (standard inference or online inference) and by different methods by considering all possible pairwise comparisons for chains corresponding to the same data set. For each pair of chains, we use the R We There Yet (RWTY) software package \citep{Warren2017} to create a plot of split frequencies as well compute their correlation and the average standard deviation of split frequencies (ASDSF) \citep{Lakner2008}. As the different chains converge to the same stationary distribution, the ASDSF should approach 0. We adopt the guideline that an ASDSF less than 0.05 (ideally, less than 0.01) supports topological convergence \citep{Ronquist2011}.

\section{Supplementary Material}

Supplementary files, tables, figures and methods will be made available online.
%are available at Molecular Biology and Evolution online (www.mbe.oxfordjournals.org/).

\section{Acknowledgements}

We would like to thank the editors, as well as Fredrik Ronquist and an anonymous reviewer for their constructive comments that helped improve this article.
The research leading to these results has received funding  from the European Research Council under the European Union's Horizon 2020 research and innovation programme (grant agreement no. 725422-ReservoirDOCS) and from the Wellcome Trust through the ARTIC Network (project 206298/Z/17/Z).
PL acknowledges support by the Special Research Fund, KU Leuven (`Bijzonder Onderzoeksfonds', KU Leuven, OT/14/115), and the Research Foundation -- Flanders (`Fonds voor Wetenschappelijk Onderzoek -- Vlaanderen', G066215N, G0D5117N and G0B9317N).
MAS is partly supported by NSF grant DMS 1264153 and NIH grants R01 AI107034 and U19 AI135995.
AR acknowledges support from the Bill \& Melinda Gates Foundation (OPP1175094 PANGEA-2).
GB acknowledges support from the Interne Fondsen KU Leuven / Internal Funds KU Leuven under grant agreement C14/18/094, and the Research Foundation -- Flanders (`Fonds voor Wetenschappelijk Onderzoek -- Vlaanderen', G0E1420N).
The computational resources and services used in this work were provided by the VSC (Flemish Supercomputer Center), funded by the Research Foundation - Flanders (FWO) and the Flemish Government -- department EWI.
We thank the many groups and individuals who generated the Ebola virus genome sequence data during the 2014-2016 West African epidemic and made them publicly available for research.

\bibliographystyle{mbe}

\end{document}